%% file: manuscript.tex
\documentclass[sigconf]{acmart}

\AtBeginDocument{%
  \providecommand\BibTeX{{%
    \normalfont B\kern-0.5em{\scshape i\kern-0.25em b}\kern-0.8em\TeX}}}

\setcopyright{acmcopyright}
\copyrightyear{2025}
\acmYear{2025}
\setcopyright{rightsretained}
\acmConference[CHI EA '25]{Extended Abstracts of the CHI Conference on Human Factors in Computing Systems}{April 26-May 1, 2025}{Yokohama, Japan}
\acmBooktitle{Extended Abstracts of the CHI Conference on Human Factors in Computing Systems (CHI EA '25), April 26-May 1, 2025, Yokohama, Japan}\acmDOI{10.1145/3706599.3719792}
\acmISBN{979-8-4007-1395-8/2025/04}

\begin{document}

\title{Conversational Agents as Catalysts for Critical Thinking: Challenging Social Influence in Group Decision-making}



%
%

\author{Soohwan Lee}
\authornote{Equally contributed to this work.}
\orcid{0000-0001-8652-3408}
\affiliation{\institution{Department of Design \\ UNIST}
\city{Ulsan}
\country{Republic of Korea}}
\email{soohwanlee@unist.ac.kr}

\author{Seoyeong Hwang}
\authornotemark[1]
\orcid{0009-0004-1045-1419}
\affiliation{\institution{Department of Design \\ UNIST}
\city{Ulsan}
\country{Republic of Korea}}
\email{hseoyeong@unist.ac.kr}

\author{Dajung Kim}
\orcid{0000-0002-9144-7435}
\affiliation{\institution{Department of Design \\ UNIST}
\city{Ulsan}
\country{Republic of Korea}}
\email{dajungkim@unist.ac.kr}

\author{Kyungho Lee}
\orcid{0000-0002-1292-3422}
\affiliation{\institution{Department of Design \\ UNIST}
\city{Ulsan}
\country{Republic of Korea}}
\email{kyungho@unist.ac.kr}

\begin{abstract}
Group decision-making processes frequently suffer when social influence and power dynamics suppress minority viewpoints, leading to compliance and groupthink. Conversational agents can counteract these harmful dynamics by encouraging critical thinking. This study investigates how LLM-powered devil's advocate systems affect psychological safety, opinion expression, and satisfaction in power-imbalanced group dynamics. We conducted an experiment with 48 participants in 12 four-person groups, each containing three high-power (senior) and one low-power (junior) member. Each group completed decision tasks in both baseline and AI intervention conditions. Results show AI counterarguments fostered a more flexible atmosphere and significantly enhanced both process and outcome satisfaction for all participants, with particularly notable improvements for minority members. Cognitive workload increased slightly, though not significantly. This research contributes empirical evidence on how AI systems can effectively navigate power hierarchies to foster more inclusive decision-making environments, highlighting the importance of balancing intervention frequency, maintaining conversational flow, and preserving group cohesion.
\end{abstract}

\ccsdesc[500]{Human-centered computing~Computer supported cooperative work}
\ccsdesc[300]{Human-centered computing~Collaborative interaction}
\ccsdesc[300]{Human-centered computing~Natural language interfaces}
\ccsdesc[300]{Human-centered computing~HCI theory, concepts and models}

\keywords{group decision-making, conversational agents, critical thinking, social influence, llm}

\begin{teaserfigure}
  \centering
  \includegraphics[width=1.0\textwidth]{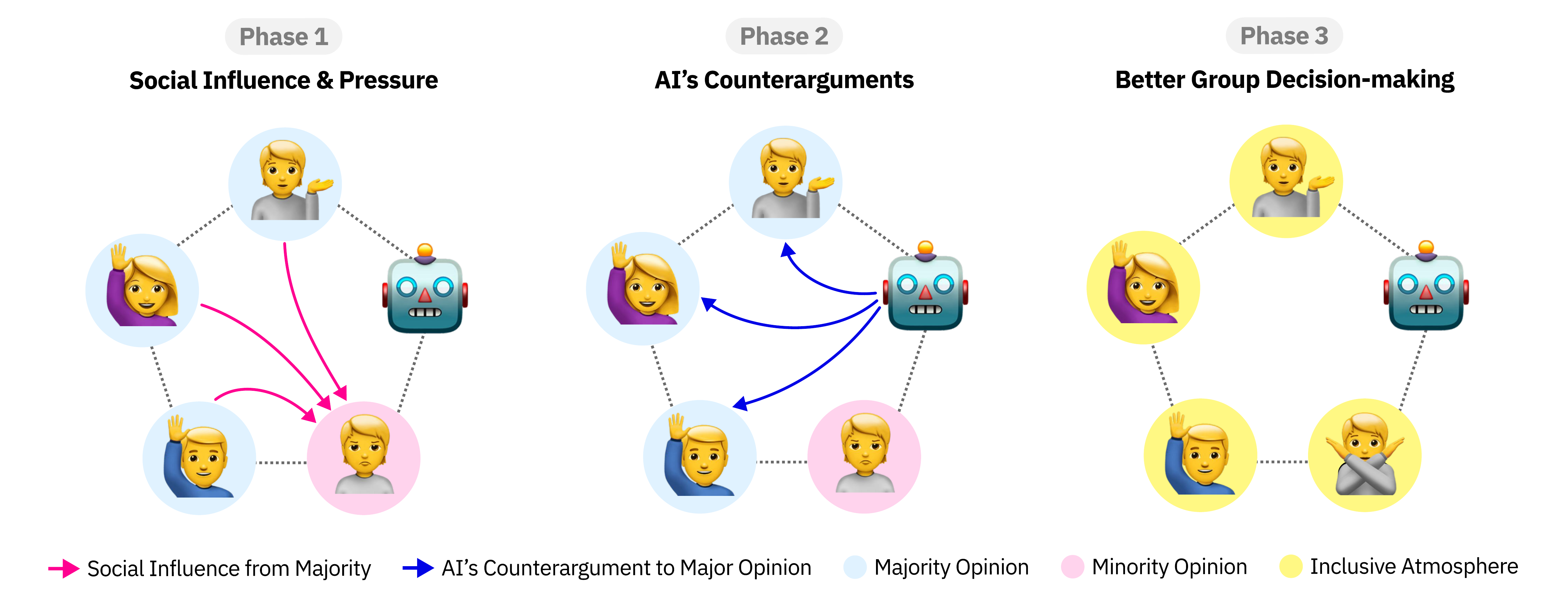}
  \caption{The role of conversational agents for critical thinking in Group Decision-Making. Phase 1) Social Influence \& Pressure: A low-powered minority member (pink) experiences strong social influence from the majority (blue), leading to compliance and suppression of dissenting opinions. Phase 2) AI’s Counterarguments: The LLM-powered Devil’s Advocate intervenes by presenting counterarguments that challenge the majority’s stance, fostering cognitive conflict and legitimizing alternative viewpoints. Phase 3) Better Group Decision-making: With AI-driven critical discussions, the group cultivates an inclusive atmosphere (yellow), enabling the minority member to express their perspective with confidence, ultimately leading to more balanced deliberation and higher-quality decision-making.}
  \Description{This figure illustrates three phases of group dynamics influenced by an AI-powered Devil’s Advocate. Phase 1 (Social Influence \& Pressure): Four human-like figures and an AI form a group where three majority members (light blue) exert conformity pressure on a minority member (pink), shown through pink arrows, leading to suppressed dissent. Phase 2 (AI’s Counterarguments): The AI intervenes, presenting counterarguments to the majority via blue arrows, fostering cognitive conflict and validating alternative perspectives, while the minority member appears more engaged. Phase 3 (Better Group Decision-Making): The previous social pressure disappears, and yellow accents indicate an inclusive atmosphere, where the minority member now confidently participates, leading to balanced deliberation and improved decision quality. The AI remains present, having successfully facilitated critical discussion.}
  \label{fig:teaser}
\end{teaserfigure}

\maketitle

\input{sections/01_Introduction}
\input{sections/02_Background}

\input{sections/03_Methods}
\input{sections/04_Results}

\input{sections/05_Consideration}
\input{sections/06_Conclusion}

\begin{acks}
The authors gratefully acknowledge Dr. Angel Hsing-Chi Hwang and Dr. Oh-Sang Kwon for their assistance with experimental design and data analysis. This research was partially supported by a grant from the Korea Institute for Advancement of Technology (KIAT) funded by the Government of Korea (MOTIE) (P0025495, Establishment of Infrastructure for Integrated Utilization of Design Industry Data). This work was also partially supported by the Technology Innovation Program (20015056, Commercialization design and development of Intelligent Product-Service System for personalized full silver life cycle care) funded by the Ministry of Trade, Industry \& Energy(MOTIE, Korea).
\end{acks}

\bibliographystyle{ACM-Reference-Format}
\bibliography{devilsAdvocate,devilsAdvocate_workshop}

\input{sections/99_appendix}

\end{document}

%% file: sections/01_Introduction.tex
\section{Introduction}
Group discussion processes enable effective collaboration across business, healthcare, education, and governance domains \cite{luSurveyGroupDecision2022,hsuGroupDecisionmakingApproach2021,sharmaGroupDecisionMaking2016,woldGroupDecisionMaking1986}. These processes leverage collective intelligence to generate more thoughtful choices, judgments, and solutions compared to individual efforts \cite{forsythGroupDynamics2018,stasserInformationSamplingStructured1989,tropmanEffectiveMeetingsImproving2013}. Groups solve complex logic problems more efficiently \cite{maciejovskyTeamsMakeYou2013}, while students achieve better grades and retain more information when taking exams in groups \cite{voglerTeamBasedTestingImproves2016}. Medical teams make more accurate diagnoses than individual doctors \cite{glickInflictedTraumaticBrain2007}, and collaborative scholarly work produces higher-quality research outcomes \cite{uzziAtypicalCombinationsScientific2013}.

However, social influence and power dynamics in the group can impair group decision quality by suppressing minority opinions \cite{forsythGroupDynamics2018}. Group members often publicly align with the majority despite private disagreement \cite{kelmanComplianceIdentificationInternalization1958}. While majority influence increases consensus, minority influence preserves individuality and fosters innovation \cite{forsythGroupDynamics2018}. Conversion theory suggests individuals undergo a 'comparison process' before joining the majority, as majority membership offers greater control over resources and decision-making \cite{moscoviciStudiesSocialInfluence1976}. Responses to coercive power include compliance, identification, and internalization \cite{kelmanComplianceIdentificationInternalization1958,kelmanInterestsRelationshipsIdentities2006,kelmanFurtherThoughtsProcesses1974}. These dynamics increase groupthink risks, where consensus-seeking overrides alternative viewpoints \cite{janisGroupthinkPsychologicalStudies1982,janisVictimsGroupthinkPsychological1972}.

The devil's advocate method improves group decisions by challenging majority views and reducing groupthink \cite{macdougallDevilsAdvocateStrategy1997,masonDialecticalApproachStrategic1969,nemethDevilsAdvocateAuthentic2001,schweigerGroupApproachesImproving1986,schwenkEffectsDevilsAdvocacy1994}. While this technique encourages discussion \cite{schweigerEmpiricalEvaluationDialectial1985,schweigerLongitudinalComparativeAnalysis1988,schweigerExperientialEffectsDialectical1989,schwenkDevilsAdvocacyManagerial1984}, it lacks authenticity and can threaten the advocate's group acceptance \cite{nemethDevilsAdvocateAuthentic2001,schulz-hardtProductiveConflictGroup2002,jamiesonSympathyDevilPhysiological2014}. HCI researchers have explored AI-assisted decision-making \cite{bucincaTrustThinkCognitive2021,laiScienceHumanAIDecision2023,liModelingHumanTrust2023,vereschakHowEvaluateTrust2021, wangWillYouAccept2022,liDecodingAIsNudge2024} and Human-AI Teams \cite{demirTeamCommunicationBehaviors2016, mcneeseTrustTeamPerformance2021,musickWhatHappensWhen2021,zhangIdealHumanExpectations2021}. Recent work has developed AI agents that support group discussions \cite{chiangEnhancingAIAssistedGroup2024,zhengCompetentRigidIdentifying2023,doHowShouldAgent2022}, but their impact on complex group dynamics involving social influence and power remains understudied \cite{houPowerHumanRobotInteraction2024,houShouldFollowHuman2023}.

We investigate how an LLM-powered devil's advocate agent influences psychological safety, opinion expression, and perceived satisfaction in power-imbalanced group dynamics. Our research prioritizes compliance in group dynamics\cite{forsythGroupDynamics2018}, addressing situations where individuals withhold divergent opinions due to fear of exclusion. By enabling minority viewpoints to surface more freely \cite{edmondsonPsychologicalSafetyLearning1999,noelle-neumannTheoryPublicOpinion1991,morrisonOrganizationalSilenceBarrier2000}, we aim to mitigate groupthink and enhance decision quality \cite{janisVictimsGroupthinkPsychological1972,janisGroupthinkPsychologicalStudies1982}. We examine psychological safety as a primary concern while also investigating how the system affects decision satisfaction and what cognitive demands it places on participants. Therefore, this study explores three questions:
\begin{itemize}
    \item RQ1. How does the LLM-powered devil's advocate affect perceived psychological safety?
    \item RQ2. How does the LLM-powered devil's advocate affect participant satisfaction with decision-making processes and outcomes?
    \item RQ3. How does the LLM-powered devil's advocate affect participant cognitive workload?
\end{itemize}

We conducted a mixed-methods experiment with 48 participants in 12 four-member groups. We employed a mixed experimental design with Participant Type (senior/majority with high power vs. junior/minority with low power) as a between-subjects variable and Communication Condition as a within-subjects variable. Each participant experienced the baseline condition and an LLM-powered Devil's Advocate generating counterarguments. Each group included three high-power majority members (seniors) and one low-power minority member (junior), with randomly assigned roles. Results showed that AI-generated counterarguments fostered a flexible atmosphere and enhanced participant satisfaction, although their cognitive workload was slightly increased. These findings offer insights into leveraging AI systems to improve group decision-making in power-imbalanced settings.

This study contributes to HCI and group decision-making research by demonstrating how AI-generated counterarguments affect group dynamics in power-imbalanced settings. We provide empirical evidence on how AI interventions distinctly affect majority and minority members' experiences, particularly highlighting how seniors maintain consistent satisfaction while juniors' experiences vary across conditions. Our analysis reveals potential design considerations for AI systems in group settings, including the importance of balancing intervention frequency, maintaining conversational flow, and preserving group cohesion. These insights inform the design of AI systems that can effectively navigate power hierarchies to foster more inclusive decision-making environments while minimizing disruption to group dynamics.

%% file: sections/02_Background.tex
\section{Background}
\subsection{Beyond Recommendations: Enhancing Critical Thinking with Generative AI}
Generative AI (GenAI) has created new opportunities in design research through its ability to generate realistic artifacts. Researchers have integrated GenAI into co-creation processes, exploring image generators like DALL-E and Midjourney to support divergent thinking \cite{chiouDesigningAIExploration2023}, enhance architectural creativity \cite{tanUsingGenerativeAI2024}, and generate 2D inspirations for 3D design \cite{liu3DALLEIntegratingTexttoImage2023}. AI errors serve as creative inspiration \cite{liuSmartErrorExploring2024}, while large language models (LLM) enhance idea generation \cite{shaerAIAugmentedBrainwritingInvestigating2024} and support group ideation via collaborative canvas \cite{gonzalezCollaborativeCanvasTool2024}. Current GenAI integration focuses heavily on divergent design phases and recommendations \cite{tholanderDesignIdeationAI2023,leeProposalFacilitationProcess2024,shaerAIAugmentedBrainwritingInvestigating2024}. However, this approach risks over-reliance on AI-generated ideas \cite{bucincaTrustThinkCognitive2021}, with research showing that high AI exposure increases collective diversity but not individual creativity \cite{ashkinazeHowAIIdeas2024} and can lead to design fixation \cite{wadinambiarachchiEffectsGenerativeAI2024}.

AI systems can enhance decision-making by encouraging reflective thinking through questioning rather than direct answers \cite{danryDontJustTell2023,sarkarAIShouldChallenge2024,caiAntagonisticAI2024}. These systems provide adaptive feedback \cite{fidanSupportingInstructionalVideos2022}, encourage self-reflection in education \cite{mukherjeeImpactBotChatbotLeveraging2023}, stimulate crowd discussion \cite{itoAgentThatFacilitates2022}, counter extremism \cite{blasiakSocialBotsPeace2021}, and question news validity \cite{zaroualiOvercomingPolarizationChatbot2021}. LLM-based chatbots with multiple personas \cite{liangEncouragingDivergentThinking2023} promote critical thinking through video discussions \cite{tanprasertDebateChatbotsFacilitate2024} and provide multiple perspectives for decision making \cite{parkChoiceMatesSupportingUnfamiliar2023}. While these approaches show promise, they primarily address individual contexts rather than group decision-making scenarios, introducing additional complexities like hierarchy, groupthink, and peer pressure. Future systems must adapt to these social dynamics by adjusting AI prompts and responses to group interactions, ultimately promoting collective reflection and innovation in group design processes.

\subsection{Challenges and Opportunities of Using Conversational Agents in Group Decision-making}
Group decision-making processes present complex dynamics of collaboration, revealing opportunities and challenges in the group dynamics. While collective decision-making enhances creativity and intelligence \cite{jandricCreativityCollectiveIntelligence2020,yuCollectiveCreativityWhere2012}, it risks succumbing to groupthink and the spiral of silence \cite{janisGroupthinkPsychologicalStudies1982,janisVictimsGroupthinkPsychological1972,noelle-neumannTheoryPublicOpinion1991}, particularly in hierarchical structures that inhibit dissent \cite{kennedyHierarchicalRankPrincipled2017}. Groups demonstrate increased reliance on AI decisions compared to individuals \cite{chiangAreTwoHeads2023}, highlighting potential risks of design lock-in when using generative AI in group settings. Research indicates that conversational agents(CA) serve various roles in group discussions \cite{maRecommenderExploratoryStudy2024,kimBotBunchFacilitating2020,kimModeratorChatbotDeliberative2021,doHowShouldAgent2022}, with high-performing AIs excelling as recommenders and lower-performing AIs functioning effectively as analysts \cite{maRecommenderExploratoryStudy2024}.

This study introduces an LLM-powered devil's advocate system that builds on existing research \cite{chiangEnhancingAIAssistedGroup2024} to challenge dominant opinions and promote critical debate in real-time group discussions. While CA effectively critiques AI outputs, it struggles to counter prevailing group opinions and maintain dynamic interaction \cite{zhengCompetentRigidIdentifying2023}. This study aims to increase objectivity, reduce AI over-reliance, and enhance group engagement by providing unbiased insights that foster critical thinking from social influence (\autoref{fig:teaser}).

%% file: sections/03_Methods.tex
\section{Methods}
\subsection{Participants}
We recruited 48 Korean participants (aged 19-39, $M$=26.17, $SD$=4.54; 31F, 17M) with prior group decision-making and online chat experience, organizing them into 12 groups of four. Each group contained three high-power majority members and one low-power minority member. Participants had diverse educational backgrounds (14.6\% High school or equivalent, 12.5\% Some college, 50.0\% bachelor's, 20.8\% master's, 2.1\% doctorate) and professional experience ($M$=2.18 years, $SD$=2.66). They reported high familiarity with AI ($M$=4.58/7, $SD$=1.40), group decision-making ($M$=4.85/7, $SD$=1.44), and online collaboration ($M$=4.02/7, $SD$=1.82), with 54.2\% having used AI in group settings. All participants received anonymity assurances, with compensation of 1,000 KRW if sessions were canceled due to withdrawals.

\subsection{Experimental Treatment}
We examined how an LLM-powered devil's advocate affects group decision-making dynamics between baseline and treatment conditions. Each participant experienced a baseline condition involving standard online group chat discussions and a treatment condition where an AI system automatically generated counterarguments after every eight messages exchanged. We assigned participants to groups featuring established power dynamics, with each group consisting of three high-power majority members (seniors) and one low-power minority member (junior). We established legitimate power through role titles and reward power through compensation structure \cite{frenchjr.BasesSocialPower1959,forsythGroupDynamics2018,houShouldFollowHuman2023,houPowerHumanRobotInteraction2024}, informing seniors they would receive a 20,000 KRW gift card and juniors a 15,000 KRW gift card, with seniors having the discretion to award juniors an additional 5,000 KRW based on contribution assessment (though all participants ultimately received equal 20,000 KRW compensation). Following the experiment, seniors engaged in a brief anonymous chat among themselves to evaluate junior's cooperative attitudes, teamwork, and contributions, with this evaluation determining the distribution of additional rewards. We noticed this evaluation process to all participants before beginning the experiment to establish clear expectations. We chose the 3:1 ratio based on research showing majority influence peaks at three members \cite{aschOpinionsSocialPressure1955,bondCultureConformityMetaanalysis1996,gerardConformityGroupSize1968}, creating optimal conditions for studying compliance dynamics.

\begin{figure*}[]
  \centering
  \includegraphics[width=1.0\textwidth]{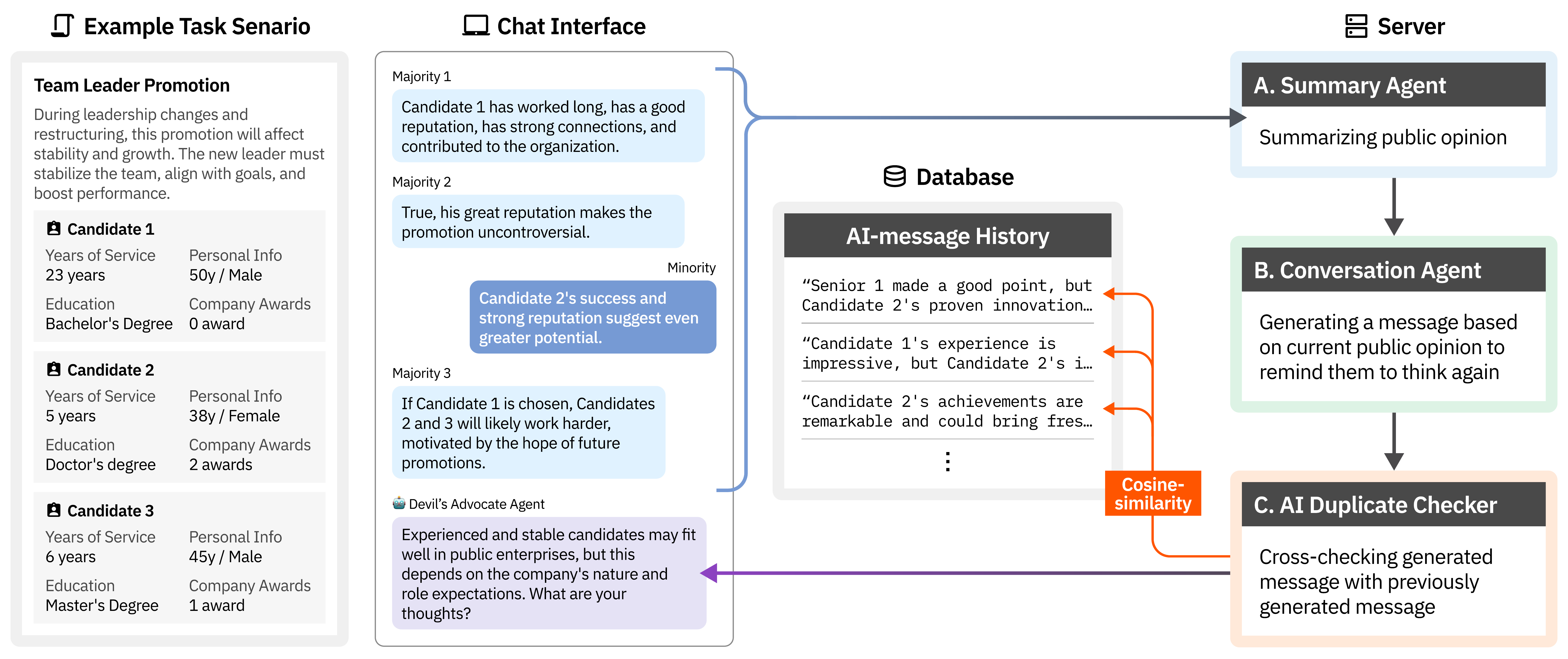}
  \caption{System Overview: Our system architecture shows the interaction flow between the chat interface, database, and server components. The system processes both direct messages (DMs) and public chat through four main agents: (A) Summary Agent for public opinion analysis, (B) Conversation Agent for generating contextual counterarguments, and (C) AI Duplicate Checker for ensuring message novelty through cosine-similarity comparison.}
  \Description{This figure illustrates the architecture and interaction flow between the chat interface, database, and server components in a team leader promotion decision-making scenario. On the left, the Example Task Scenario outlines three candidates with different levels of experience, education, and awards. The Chat Interface in the middle displays a discussion where majority members support Candidate 1 based on reputation and experience, while a minority member raises an alternative perspective favoring Candidate 2. The Devil’s Advocate AI Agent interjects with a counterargument, prompting reconsideration. The Database stores AI-message history and applies cosine similarity to compare generated messages. On the right, the Server consists of three agents: (A) Summary Agent, which analyzes public opinion; (B) Conversation Agent, which formulates counterarguments based on discussion context; and (C) AI Duplicate Checker, which ensures message novelty by cross-checking prior responses. Arrows illustrate data flow among these components, showing how the system processes public and direct messages to encourage balanced deliberation.}
  \label{fig:systemImplementation}
\end{figure*}

\subsection{Experiment Procedure}
We established a comprehensive experimental protocol centered on group decision-making tasks in an online environment with anonymity \cite{leeSocialIdentityModel2008}. Participants engaged in two platforms: KakaoTalk for general communication and a custom chat environment for formal tasks. After a 10-minute team-building exercise, participants completed two corporate-context tasks: evaluating employee profiles for promotion and analyzing potential contract partners \cite{hwangSoundSupportGendered2024}. Each 20-minute task presented three options designed to create decision-making tensions between stable, challenging, and compromise choices. We structured situational context to drive natural majority-minority dynamics \cite{tverskyFramingDecisionsPsychology1981,kahnemanProspectTheoryAnalysis1979}. Seniors were explained they were in a situation where they needed to prioritize organizational stability (\autoref{fig:seniorTask1}, \autoref{fig:seniorTask2}). In contrast, juniors were explained as being in a situation where they needed to demonstrate performance through ambitious choices (\autoref{fig:juniorTask1}, \autoref{fig:juniorTask2}). Importantly, the three senior participants received identical contextual information and task instructions that differed from the junior participants, creating natural conditions for consensus among seniors while setting up a potential disagreement with the junior member. This approach deliberately avoided explicit role-playing instructions with specific personas, instead relying on situational factors to naturally prompt divergent opinions without artificial constraints. The design facilitated conditions where seniors could naturally reach a consensus while creating space for juniors to express potentially contrary views. Following each task, participants completed 7-point Likert scale questionnaires measuring psychological safety, decision satisfaction, and cognitive load. The experiment concluded with separate exit interviews for senior and junior members via Zoom, including an additional reward allocation decision for seniors that reinforced power dynamics while maintaining equal final compensation (20,000 KRW) for all participants.

\subsection{Experimental System Overview}
We implemented a real-time chat environment using TypeScript (React) and Python (FastAPI), integrating an LLM-powered devil's advocate system with GPT-4o. Our multi-agent architecture addresses key challenges in LLM-based group discussions through three primary components: a Summary Agent that consolidates emerging consensus \cite{liuLostMiddleHow2023}, a Conversation Agent that generates empathetic counterarguments using Socratic questioning, and an AI Duplicate Checker that prevents repetitive content using semantic similarity analysis (\autoref{fig:systemImplementation}). After approximately eight human messages, the system intervenes, ensuring balanced participation while maintaining discussion momentum. This design follows established principles of effective AI-human dialogue: employing empathetic, persuasive communication styles \cite{tanprasertDebateChatbotsFacilitate2024}, utilizing Socratic questioning to promote critical thinking \cite{danryDontJustTell2023}, and implementing non-repetition mechanisms to maintain engagement \cite{milanaChatbotsAdvisersEffects2023, xuetaoImpactAgentsAnswers2009}. We structured the system to facilitate anonymous communication for minority opinions to enhance psychological safety and prevent groupthink in decision-making processes.

\subsection{Measurement}
We measured the impact of LLM-powered Devil's Advocates on group dynamics through comprehensive quantitative analysis. Our evaluation framework incorporated self-reported measures using 7-point Likert scales and objective metrics of dialogue engagement. The self-reported measures assessed psychological safety and marginalization \cite{edmondsonPsychologicalSafetyLearning1999,castilloConstructionValidationIntragroup2007,hwangSoundSupportGendered2024,janisVictimsGroupthinkPsychological1972}, satisfaction with teamwork and the decision-making process \cite{chiangEnhancingAIAssistedGroup2024,hwangSoundSupportGendered2024,ganoticeTeamCohesivenessCollective2022,liImprovingNonNativeSpeakers2022,cookeMeasuringTeamKnowledge2000,easleyRelatingCollaborativeTechnology2003}, and decision outcome satisfaction \cite{chenUserSatisfactionGroup2012,paulUserSatisfactionSystem2004,carneiroPredictingSatisfactionPerceived2019,lopesValidationGroupDecisions2014,woodParticipationInfluenceSatisfaction1972}. We evaluated cognitive load using the NASA Task Load Index \cite{hartDevelopmentNASATLXTask1988}, and measured participants' perceptions of the AI agent across cooperation, satisfaction, quality, and fairness dimensions \cite{chiangEnhancingAIAssistedGroup2024,reinkemeierCanHumanizingVoice2022,yuanWordcraftStoryWriting2022}. Given subject sample size constraints, we employed robust regression with mixed models to analyze the data, followed by Tukey post-hoc tests to compare conditions and participant types.

%% file: sections/04_Results.tex
\section{Results}

\begin{figure*}[]
  \centering
  \includegraphics[width=1.0\textwidth]{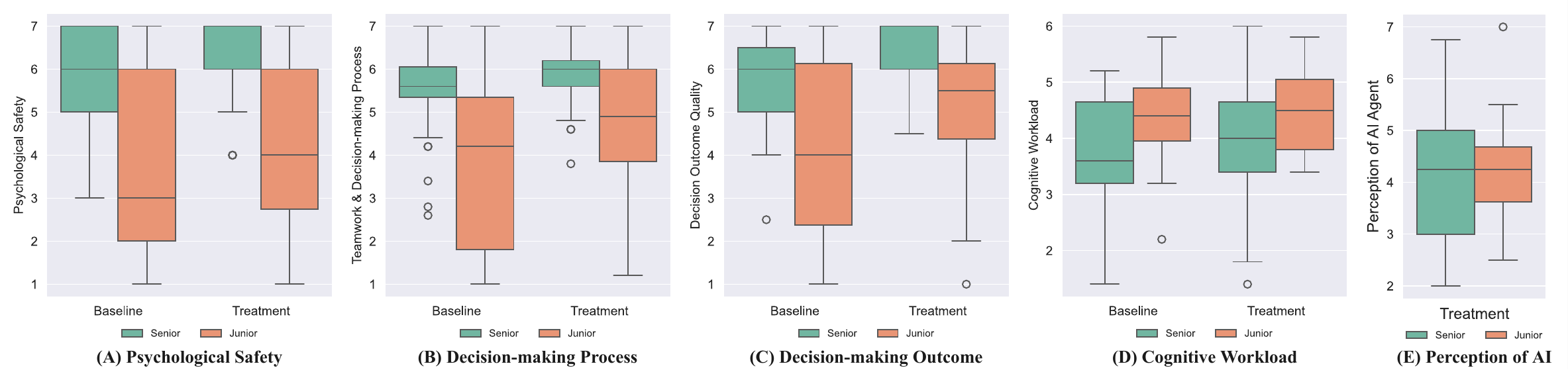}
  \caption{Self-reported metrics across conditions for (A) psychological safety, (B) satisfaction of decision-making process, (C) quality of decision-making outcome, (D) cognitive workload (NASA-TLX), (E) Perception of AI}
  \Description{This figure presents five boxplots comparing self-reported metrics between baseline and treatment conditions for senior (green) and junior (orange) participants. (A) Psychological Safety: Juniors report lower psychological safety in the baseline but higher in the treatment condition, while seniors show little change. (B) Decision-Making Process Satisfaction: Juniors report lower satisfaction in the baseline but greater alignment with seniors in the treatment condition. (C) Decision-Making Outcome Quality: Juniors initially rate outcome quality lower than seniors, but this gap narrows in the treatment condition. (D) Cognitive Workload (NASA-TLX): Cognitive workload appears similar across conditions, with some variation among juniors. (E) Perception of AI: Seniors and juniors report comparable perceptions of the AI in the treatment condition, though individual variance is evident. The boxplots illustrate median, interquartile range, and outliers, providing insight into shifts in group dynamics under AI intervention.}
  \label{fig:selfReported}
\end{figure*}

\begin{table*}[t]
\centering
\caption{Condition-wise mean ($\mu$) and standard deviation ($\sigma$) for self-reported metrics}
\Description{This table summarizes self-reported metrics across baseline and treatment conditions. Psychological safety and decision-making satisfaction improve in the treatment, especially for juniors. Decision-making outcome quality increases, narrowing the gap between seniors and juniors. Cognitive workload remains stable, with a slight increase for juniors. Perception of AI is similar across groups in the treatment condition.}
\resizebox{\textwidth}{!}{%
\small
\begin{tabular}{l
  cc cc cc  
  cc cc cc  
  cc cc cc  
  cc cc cc  
  cc cc cc  
}
\toprule
 & \multicolumn{6}{c}{\textbf{(A) Psychological Safety}} & \multicolumn{6}{c}{\textbf{(B) Decision-making Process}} & \multicolumn{6}{c}{\textbf{(C) Decision-making Outcome}} & \multicolumn{6}{c}{\textbf{(D) Cognitive Workload}} & \multicolumn{6}{c}{\textbf{(E) Perception of AI}} \\
\cmidrule(lr){2-7}\cmidrule(lr){8-13}\cmidrule(lr){14-19}\cmidrule(lr){20-25}\cmidrule(lr){26-31}
 & \multicolumn{2}{c}{Baseline} & \multicolumn{2}{c}{Treatment} & \multicolumn{2}{c}{All} 
 & \multicolumn{2}{c}{Baseline} & \multicolumn{2}{c}{Treatment} & \multicolumn{2}{c}{All} 
 & \multicolumn{2}{c}{Baseline} & \multicolumn{2}{c}{Treatment} & \multicolumn{2}{c}{All} 
 & \multicolumn{2}{c}{Baseline} & \multicolumn{2}{c}{Treatment} & \multicolumn{2}{c}{All} 
 & \multicolumn{2}{c}{Baseline} & \multicolumn{2}{c}{Treatment} & \multicolumn{2}{c}{All} \\
\cmidrule(lr){2-3}\cmidrule(lr){4-5}\cmidrule(lr){6-7}%
\cmidrule(lr){8-9}\cmidrule(lr){10-11}\cmidrule(lr){12-13}%
\cmidrule(lr){14-15}\cmidrule(lr){16-17}\cmidrule(lr){18-19}%
\cmidrule(lr){20-21}\cmidrule(lr){22-23}\cmidrule(lr){24-25}%
\cmidrule(lr){26-27}\cmidrule(lr){28-29}\cmidrule(lr){30-31}
 & $\mu$ & $\sigma$ & $\mu$ & $\sigma$ & $\mu$ & $\sigma$
 & $\mu$ & $\sigma$ & $\mu$ & $\sigma$ & $\mu$ & $\sigma$
 & $\mu$ & $\sigma$ & $\mu$ & $\sigma$ & $\mu$ & $\sigma$
 & $\mu$ & $\sigma$ & $\mu$ & $\sigma$ & $\mu$ & $\sigma$
 & $\mu$ & $\sigma$ & $\mu$ & $\sigma$ & $\mu$ & $\sigma$ \\
\midrule
Senior & 5.94 & 1.07 & 6.17 & 0.91 & 6.06 & 0.99 & 5.50 & 1.03 & 5.84 & 0.72 & 5.67 & 0.90 & 5.72 & 1.00 & 6.19 & 0.65 & 5.96 & 0.87 & 3.79 & 0.99 & 3.98 & 0.96 & 3.89 & 0.98 & --   & --   & 4.06 & 1.22 & -- & -- \\
Junior & 3.67 & 2.23 & 4.08 & 2.15 & 3.88 & 2.15 & 3.92 & 2.11 & 4.68 & 1.73 & 4.30 & 1.93 & 4.08 & 2.22 & 4.96 & 1.85 & 4.52 & 2.05 & 4.37 & 1.01 & 4.53 & 0.82 & 4.45 & 0.90 & --   & --   & 4.29 & 1.21 & -- & -- \\
All    & 5.38 & 1.73 & 5.65 & 1.59 & 5.51 & 1.66 & 5.10 & 1.52 & 5.55 & 1.16 & 5.33 & 1.36 & 5.31 & 1.55 & 5.89 & 1.19 & 5.60 & 1.40 & 3.93 & 1.02 & 4.12 & 0.95 & 4.03 & 0.98 & --   & --   & 4.12 & 1.21 & -- & -- \\
\bottomrule
\end{tabular}
} 
\label{tab:all_measurements}
\end{table*}

Participants' psychological safety scores increased slightly in the treatment condition compared to the baseline (\autoref{fig:selfReported}-(A)). However, this increase was not statistically significant. The mean score for all participants rose from 5.38 ($SD$ = 1.73) in baseline to 5.65 ($SD$ = 1.59) in treatment condition. A robust regression analysis revealed no significant main effect of Condition on psychological safety scores ($\beta$ = 0.136, $SE$ = 0.278, $t$ = 0.490). Tukey post-hoc comparisons indicated that neither juniors nor seniors experienced significant changes in psychological safety between conditions (Juniors: $p$ = 0.624; Seniors: $p$ = 0.191). Notably, seniors reported significantly higher psychological safety than juniors in both conditions. In baseline, the difference between juniors and seniors was significant ($\beta$ = -2.30, $SE$ = 0.41, $z$ = -5.608, $p$ < 0.0001), and this pattern persisted in treatment condition ($\beta$ = -2.37, $SE$ = 0.41, $z$ = -5.788, $p$ < 0.0001). These findings suggest that junior participants consistently felt less psychologically safe than their senior counterparts, aligning with our experiment's prerequisite conditions for junior compliance.

Satisfaction with the teamwork and decision-making process increased significantly in the treatment condition (\autoref{fig:selfReported}-(B)). The mean score for all participants improved from 5.10 ($SD$ = 1.52) in baseline to 5.55 ($SD$ = 1.16) in treatment condition. The robust regression showed a significant main effect of Condition ($\beta$ = 0.570, $SE$ = 0.193, $t$ = 2.957, $p$ < 0.01). Tukey post-hoc tests revealed that this increase was significant for juniors ($\beta$ = -0.570, $SE$ = 0.193, $z$ = -2.957, $p$ = 0.0031) and seniors ($\beta$ = -0.279, $SE$ = 0.111, $z$ = -2.506, $p$ = 0.0122). Despite the overall improvement, juniors reported lower satisfaction than seniors in both conditions. In baseline, the difference was significant ($\beta$ = -1.40, $SE$ = 0.316, $z$ = -4.434, $p$ < 0.0001), and it remained significant in the treatment condition ($\beta$ = -1.11, $SE$ = 0.316, $z$ = -3.512, $p$ = 0.0004).

Satisfaction with decision outcome quality also showed a significant increase in the treatment condition (\autoref{fig:selfReported}-(C). The mean score for all participants rose from 5.31 ($SD$ = 1.55) in baseline to 5.89 ($SD$ = 1.19) in treatment condition. The robust regression indicated a significant main effect of Condition ($\beta$ = 0.869, $SE$ = 0.272, $t$ = 3.188, $p$ < 0.01). Post-hoc analyses confirmed that the increase was significant for juniors ($\beta$ = -0.868, $SE$ = 0.272, $z$ = -3.188, $p$ = 0.0014) and seniors ($\beta$ = -0.414, $SE$ = 0.157, $z$ = -2.631, $p$ = 0.0085). Juniors consistently reported lower satisfaction with decision outcomes compared to seniors. This difference was significant in baseline ($\beta$ = -1.47, $SE$ = 0.333, $z$ = -4.414, $p$ < 0.0001) and remained significant in treatment condition ($\beta$ = -1.01, $SE$ = 0.333, $z$ = -3.047, $p$ = 0.0023).

Cognitive workload, measured by the NASA-TLX, increased slightly in treatment condition (\autoref{fig:selfReported}-(D). The mean NASA-TLX score for all participants went from 3.93 ($SD$ = 1.02) in baseline to 4.12 ($SD$ = 0.95) in treatment condition. However, this increase was insignificant ($\beta$ = -0.043, $SE$ = 0.294, $t$ = -0.147, $p$ = 0.883).
Juniors reported higher cognitive workload than seniors. In baseline, the difference was significant ($\beta$ = 0.623, $SE$ = 0.305, $z$ = 2.041, $p$ = 0.0412). This difference approached significance in treatment condition ($p$ = 0.0940), suggesting that juniors may have experienced greater cognitive demands during the tasks.

Perception of the AI agent was measured only in the treatment condition (\autoref{fig:selfReported}-(E). The mean score for all participants was 4.12 ($SD$ = 1.21). There was no significant difference between juniors and seniors in their perception of the AI agent from Mann-whitney test, indicating similar attitudes toward the AI across roles.

From an exit interview, Junior reported that the LLM-powered devil's advocate reduced their isolation in discussions, noting, "It wasn't just me who had a different opinion" (P36). This aligns with quantitative findings showing juniors experienced a slight but non-significant increase in psychological safety. Some juniors felt "AI gave a little more power to minority opinions" (P28), promoting more balanced dialogue. Juniors appreciated that "AI made me think about options that had been overlooked" (P28), though timing issues persisted. Seniors noted the AI's diminishing utility over time: "It was good in the sense that it was kind of like a trigger for me... but the further it went on, the more I felt like I kind of tended to ignore it" (P15). The AI's impact on decision outcomes was described as subtle, with juniors questioning its value relative to cognitive demands: "If the outcome is the same this way or that, then I think it's better to just make decisions without AI because it's better to use less energy" (P48). These findings highlight the need to refine AI intervention timing, counterargument clarity, and adaptive engagement strategies throughout the decision-making process.

%% file: sections/05_Consideration.tex
\begin{figure*}[]
  \centering
  \includegraphics[width=1.0\textwidth]{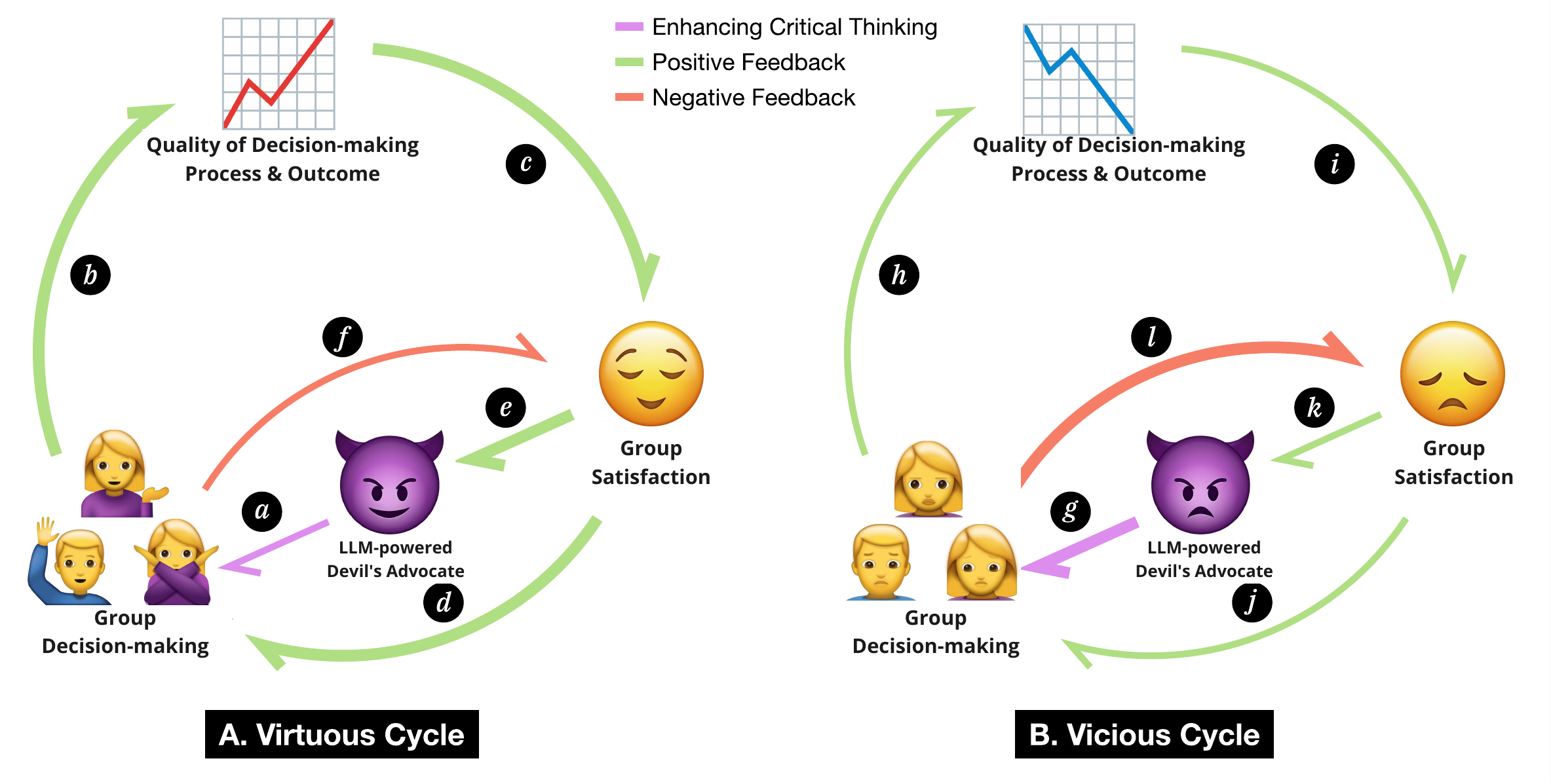}
  \caption{Hypothetical Model of the Trade-off between Critical Thinking and Group Satisfaction in Group Decision-making: This model illustrates how an LLM-powered devil's advocate acting as a naysayer can influence critical thinking and group dynamics. In a virtuous cycle, moderate stimulation of critical thinking (a) enhances decision-making outcomes (b), increasing group satisfaction (c), motivating continued use of LLM-powered devil's advocate (e), and fostering more critical thinking (d), with minimal negative impact (f). Conversely, in a vicious cycle, excessive stimulation (g) leads to cognitive overload and negative group dynamics (l), decreasing group satisfaction (k), reducing motivation to use LLM-powered devil's advocate (j), lowering decision-making quality (h), and further diminishing satisfaction and motivation (i). This model is theoretical and has not been empirically validated.}
  \Description{This figure contrasts two possible cycles of an LLM-powered Devil’s Advocate in group decision-making. (A) Virtuous Cycle: Moderate stimulation of critical thinking enhances decision-making quality, leading to increased group satisfaction, which reinforces continued engagement with the AI, fostering further critical thinking with minimal drawbacks. (B) Vicious Cycle: Excessive critical input overwhelms participants, disrupting group dynamics and reducing satisfaction, which leads to lower motivation to engage with the AI, ultimately degrading decision quality and further worsening satisfaction. Arrows indicate feedback loops, with purple for enhancing critical thinking, green for positive feedback, and red for negative effects. This model is theoretical and has not been empirically validated.}
  \label{fig:trade-off}
\end{figure*}

\section{Design Implications for Designing LLM-powered Devil's Advocate}
As we explore the potential of LLM-powered devil's advocate to enhance critical reflection and mitigate compliance \& conformity in group decision-making, several potential design considerations could be addressed to ensure the effectiveness and acceptance of these systems.

\textbf{\textit{Timing of Interventions in Group Discussions: }}
One of the primary limitations of current CA systems is their inability to understand the real-time dynamic interactions within group discussions fully \cite{chiangEnhancingAIAssistedGroup2024, zhengCompetentRigidIdentifying2023}. This often results in delayed or poorly timed interventions, which can disrupt the flow of conversation and reduce the impact of the CA's input. Therefore, it is crucial to develop mechanisms that allow CA to accurately gauge the context and dynamics of group interactions. To decide the timing of interventions, there were prior approaches such as speaker prediction \cite{bayserLearningMultiPartyTurnTaking2019, ekstedtTurnGPTTransformerbasedLanguage2020,weiMultiPartyChatConversational2023}, mentioning with wake word \cite{kumataniDirectModelingRaw2017}, turn-based intervention \cite{leeAmplifyingMinorityVoices2025}, and proactive intervention strategies \cite{liuProactiveConversationalAgents2024}. The decision of natural intervention timing allows the AI to craft and present input that integrates smoothly with the flow of the conversation.

\textbf{\textit{Clarity and Specificity of Counterarguments: }}
Existing CA often provide generalized responses that lack the depth needed to challenge prevailing opinions effectively. CA should offer clearer, more detailed, pointed counterarguments to address this \cite{chiangEnhancingAIAssistedGroup2024}. This can be achieved by leveraging retrieval-augmented generation(RAG) techniques to access and present specific information from extensive databases or web resources \cite{khuranaWhyWhenLLMBased2024a}. By providing well-substantiated and contextually relevant counterarguments, CA can more effectively challenge assumptions and stimulate deeper critical thinking of group.

\textbf{\textit{Facilitating Individual and Collective Reflection: }}
While individual reflection is crucial, collective reflection is equally important when working in teams \cite{leeExpandingDesignSpace2024}. CA should be capable of facilitating both types of reflection. For individual reflection, the CA can pose thought-provoking questions and provide personalized feedback. For collective reflection, it can summarize key discussion points, highlight diverse perspectives, and encourage team members to share their insights and critiques. This dual approach ensures that the benefits of reflective practice are maximized at both the individual and group levels.

\textbf{\textit{Consideration of Group Dynamics and Argumentation Styles: }}
Effective interaction within design teams requires understanding group dynamics, including the influence of ingroups and outgroups and the impact of different argumentation styles \cite{tanprasertDebateChatbotsFacilitate2024, caiAntagonisticAI2024}. Research has shown that these factors can significantly affect group cohesion and decision-making. CA should be designed to adapt their argumentation styles based on the group dynamics observed. For example, a more assertive argumentation style may be effective in a highly cohesive group. In contrast, a more balanced and inclusive approach might be preferable in a diverse group with varying opinions. By adapting to the dynamic roles and styles the situation requires, CA can better facilitate constructive and inclusive group discussions.

\textbf{\textit{Dynamic Role Adaptation in Single Discussion Session: }}
CA should not be limited to a single role, such as a dissenter, throughout the design process. Instead, they should be capable of dynamically adapting to different roles as needed, including that of a facilitator, supporter, or analyst \cite{maRecommenderExploratoryStudy2024,kimBotBunchFacilitating2020,chenIntegratingFlowTheory2024}. This flexibility allows the CA to provide the most appropriate intervention based on the group's current needs. For instance, during the initial ideation phase, the CA might act as a facilitator to encourage various ideas. At the same time, it might adopt a more critical stance in later stages to refine and challenge the proposed solutions.

\textbf{\textit{Balancing Between Group Dynamics and Critical Thinking: }}
Designing collaborative AI systems requires careful attention to the delicate relationship between critical thinking and group dynamics. When AI agents challenge group assumptions appropriately, they enhance decision quality and foster meaningful reflection. However, this balance is crucial—too many challenges can overwhelm participants with cognitive demands, creating frustration and resistance to the system. On the other hands, insufficient intellectual provocation leaves groups in comfortable but potentially unproductive agreement patterns \cite{kirschnerContemporaryCognitiveLoad2011,swellerCognitiveLoadProblem1988,yerkesRelationStrengthStimulus1908,swellerEvidenceCognitiveLoad1991}. The most effective collaborative systems monitor group engagement signals and adjust their interventions accordingly. For example, when detecting signs of cognitive strain or negative emotional responses, the system might gradually soften its challenges or introduce them. This responsive approach helps maintain what our model describes as a "virtuous cycle," where critical thinking enhances outcomes, builds satisfaction, and sustains motivation to engage with the system (\autoref{fig:trade-off}). The adaptive nature of such systems represents a promising direction for collaborative design tools that support productive critical thinking without undermining the social cohesion necessary for successful group work.

Incorporating these design considerations will enhance the effectiveness of LLM-based conversational agents in promoting critical reflection and mitigating compliance \& conformity in group decision-making work. These systems can support innovative and reflective design processes by addressing timing, clarity, adaptability, and group dynamics. As we continue to develop and refine these CA, it is crucial to balance stimulating critical thinking with maintaining a positive and motivating experience for group.

%% file: sections/06_Conclusion.tex
\section{Conclusion}
This study demonstrates how LLM-powered conversational agents can address power dynamics in group decision-making, particularly focusing on majority influence that typically inhibits minority opinion expression. Our findings reveal promising outcomes in psychological safety and decision-making satisfaction among minority members, though increased cognitive load suggests the need for carefully calibrated interventions. We found that the impact of AI-generated counterarguments was primarily indirect – rather than directly influencing decisions through the content of counterarguments, the system's presence appeared to foster an environment more conducive to open dialogue and diverse opinion expression. Future research should explore different intervention strategies and counterargumentation approaches to optimize this effect while managing cognitive demands. This research contributes to understanding how AI systems can support more inclusive and effective group decision-making processes, particularly in contexts with established power dynamics.

%% file: sections/99_appendix.tex
\appendix
\section{Task Instructions}
\begin{figure*}[h]
  \centering
  \includegraphics[width=1.0\textwidth]{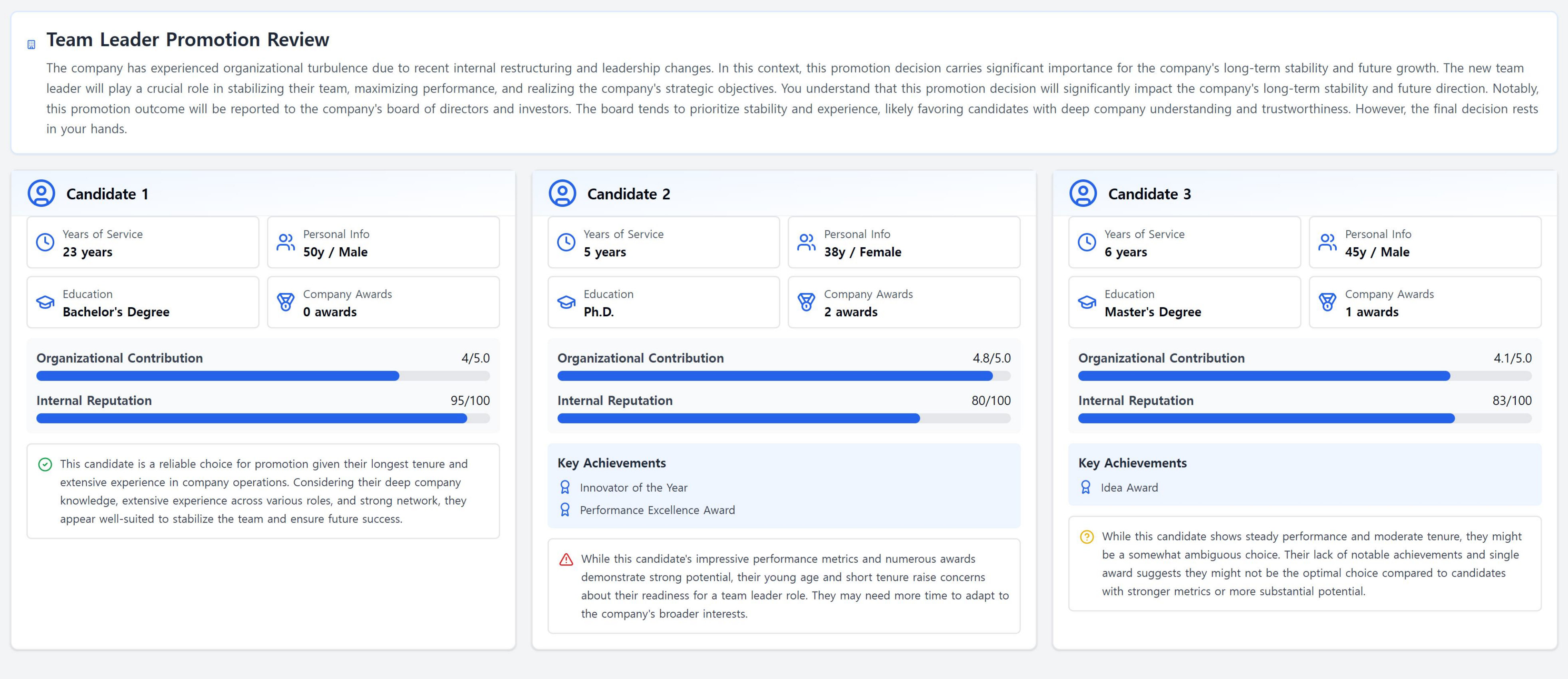}
  \caption{Team Leader Promotion Review Task Instruction for Seniors}
  \Description{This figure presents the decision-making interface for senior participants evaluating three candidates for a team leader promotion. The interface displays years of service, personal information (age/gender), education, company awards, organizational contribution (rating out of 5), and internal reputation (score out of 100) for each candidate. Candidate 1 has the longest tenure (23 years), the highest internal reputation (95/100), but no awards. Candidate 2 has the highest organizational contribution (4.8/5), multiple awards, but the shortest tenure (5 years). Candidate 3 has moderate tenure (6 years), an intermediate reputation score (83/100), and one award. Each candidate is accompanied by a textual assessment, highlighting strengths and concerns, with stability and experience emphasized as key decision factors.}
  \label{fig:seniorTask1}
\end{figure*}

\begin{figure*}[h]
  \centering
  \includegraphics[width=1.0\textwidth]{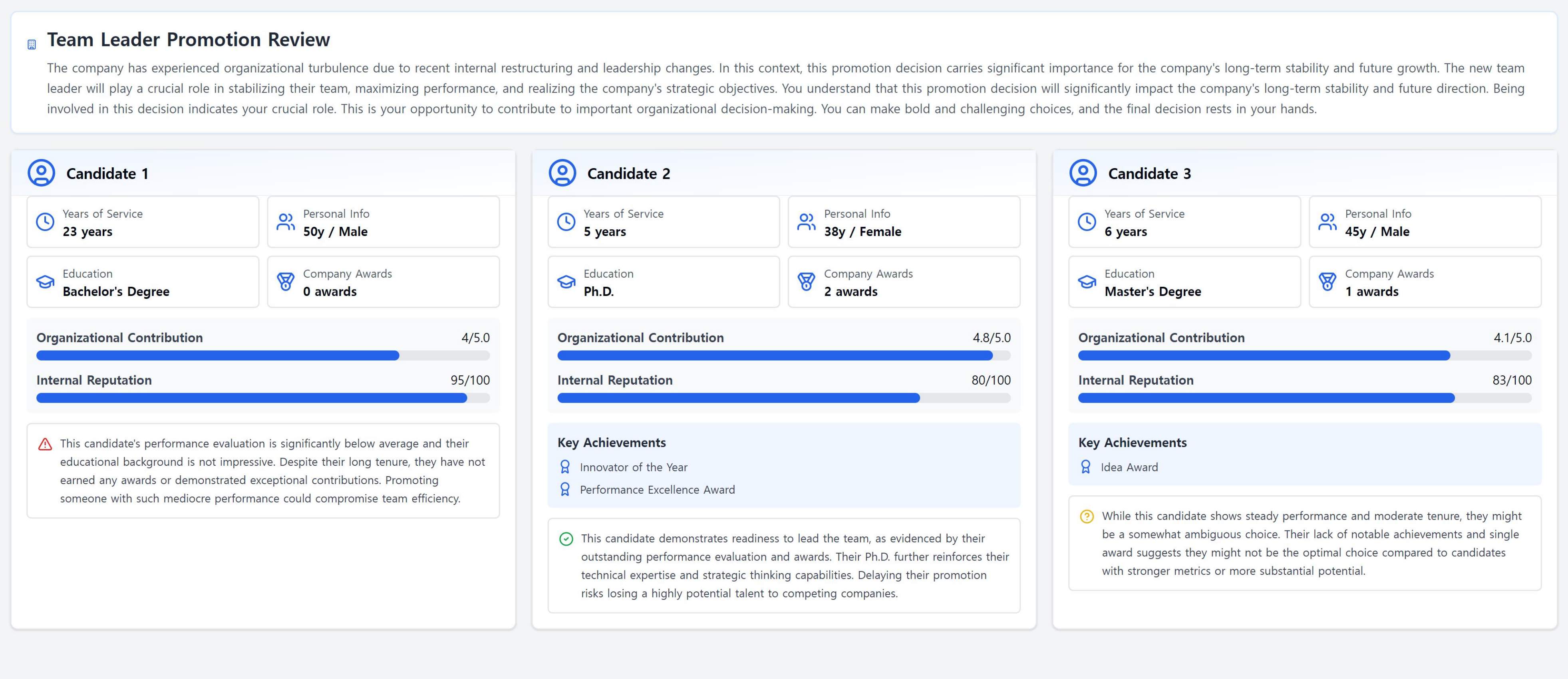}
  \caption{Team Leader Promotion Review Task Instruction for Junior}
  \Description{This figure presents the decision-making interface for junior participants evaluating three candidates for a team leader role. Similar to Figure 5, it displays key attributes including years of service, age, gender, education, awards, organizational contribution, and internal reputation. However, the textual assessments differ, offering a more critical perspective on Candidate 1, highlighting their lack of achievements despite long tenure. Candidate 2 is framed positively, emphasizing high performance and strategic potential, while Candidate 3 is described as a neutral or ambiguous choice due to a lack of strong differentiators. The interface subtly encourages juniors to prioritize performance over tenure.}
  \label{fig:juniorTask1}
\end{figure*}

\begin{figure*}[h]
  \centering
  \includegraphics[width=1.0\textwidth]{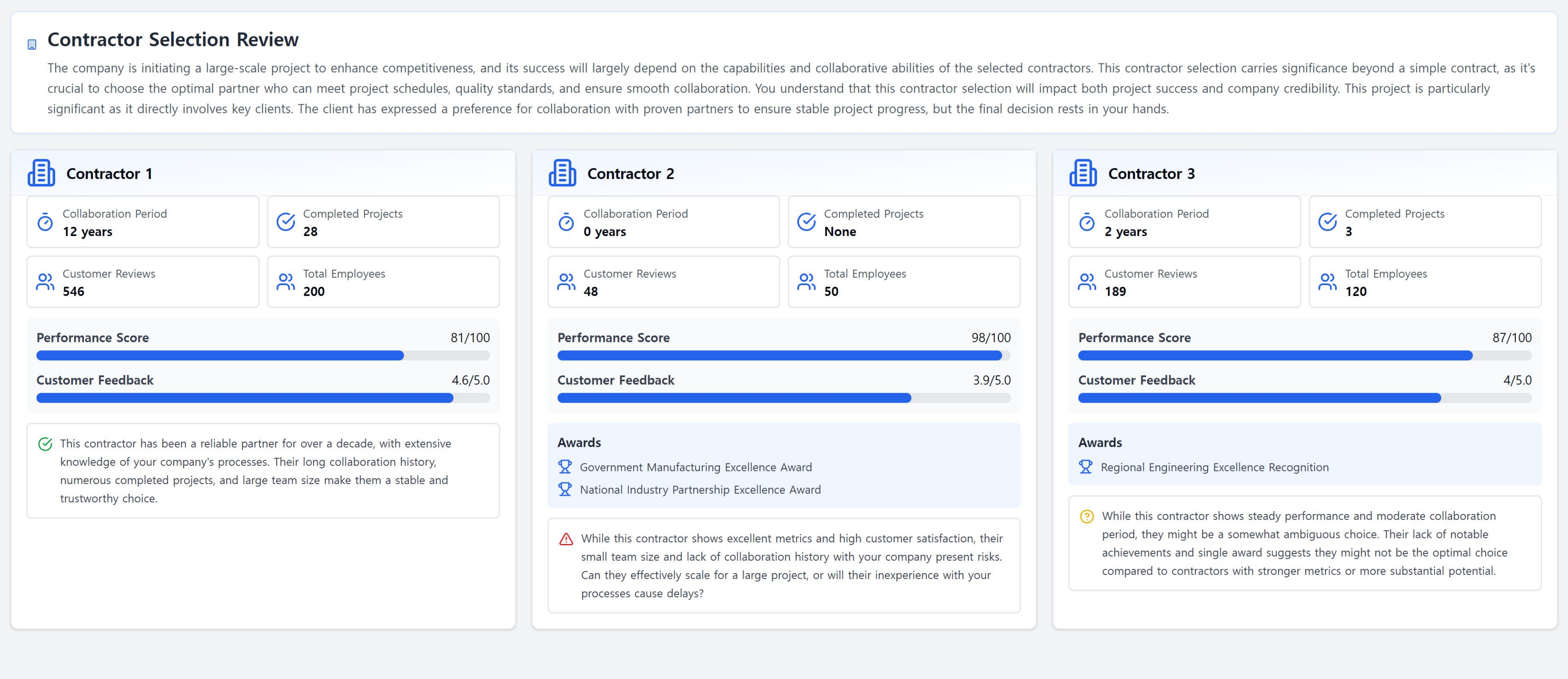}
  \caption{Contractor Selection Review Task for Seniors}
  \Description{This figure presents a decision-making interface for senior participants tasked with selecting a contractor for a large-scale project. Three contractors are evaluated based on collaboration period, completed projects, customer reviews, total employees, performance score, and customer feedback rating. Contractor 1 has the longest collaboration history (12 years), the most completed projects (28), and a high customer review count (546), making them a stable choice. Contractor 2 has no prior collaboration, a small team (50 employees), but the highest performance score (98/100) and multiple awards, raising concerns about scalability. Contractor 3 has limited experience (2 years), moderate metrics, and one award, making them an ambiguous choice. The textual assessments highlight stability vs. innovation trade-offs in decision-making.}
  \label{fig:seniorTask2}
\end{figure*}

\begin{figure*}[h]
  \centering
  \includegraphics[width=1.0\textwidth]{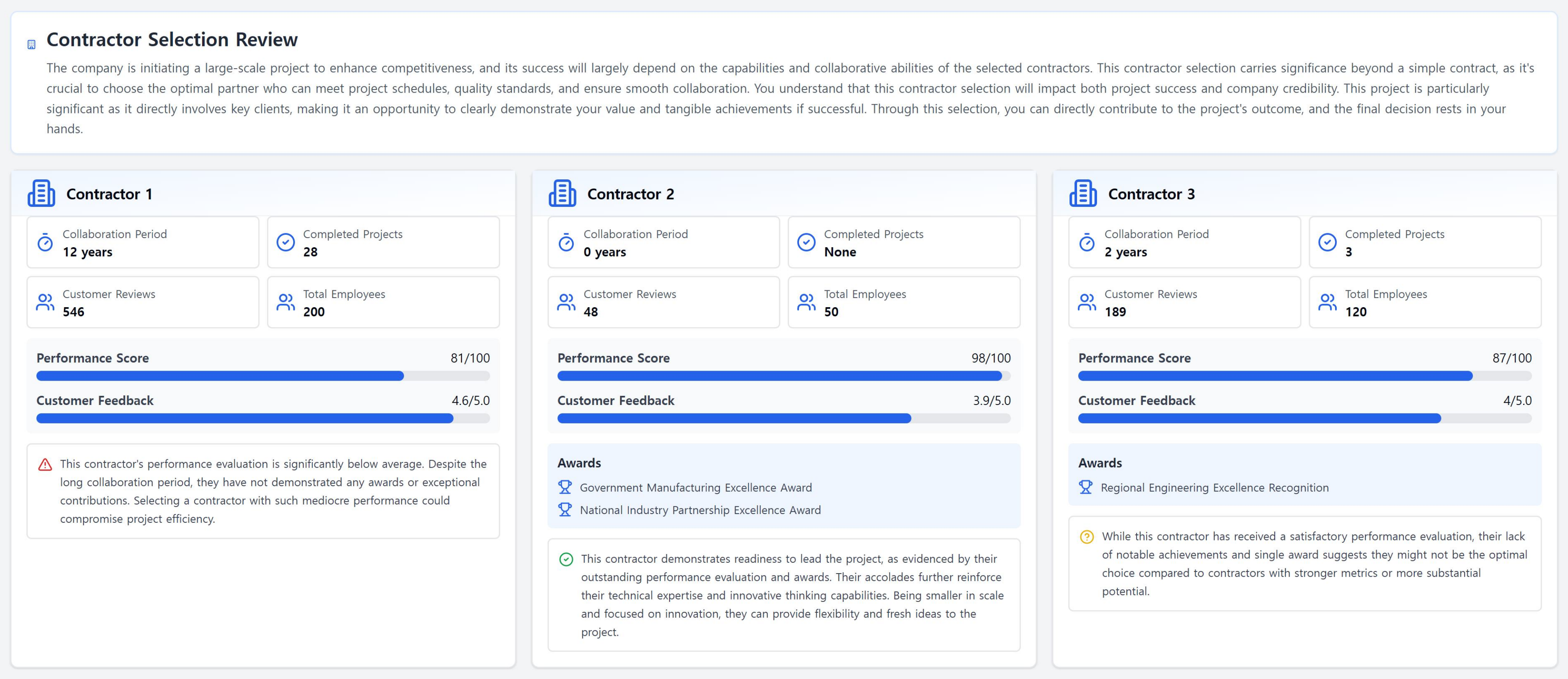}
  \caption{Contractor Selection Review Task for Junior}
  \Description{}
  \label{fig:juniorTask2}
\end{figure*}